%% file: main.tex
\pdfoutput = 1
\documentclass[conference]{IEEEtran}
\IEEEoverridecommandlockouts
\usepackage{hyperref}
\usepackage{cite}
\usepackage{amsmath,amssymb,amsfonts}
\usepackage{algorithmic}
\usepackage{graphicx}
\usepackage{textcomp}
\usepackage{float}
\usepackage{xcolor}
\usepackage[a4paper, total={184mm,239mm}]{geometry}
\def\BibTeX{{\rm B\kern-.05em{\sc i\kern-.025em b}\kern-.08em
    T\kern-.1667em\lower.7ex\hbox{E}\kern-.125emX}}
\usepackage{siunitx}
\DeclareSIUnit{\nothing}{\relax}
\usepackage{multirow}
\usepackage{booktabs}
\DeclareSIUnit{\nothing}{\relax}

\usepackage{xcolor}

\begin{document}

\title{
\textcolor{gray}{\Large This paper has been accepted for pubblication in the IEEE ISCAS2023 conference.} \\
Cyber Security aboard Micro Aerial Vehicles: An OpenTitan-based Visual Communication Use Case
}

\author{\IEEEauthorblockN{
 Maicol Ciani\IEEEauthorrefmark{1}, Stefano Bonato\IEEEauthorrefmark{2}, Rafail Psiakis\IEEEauthorrefmark{3}, Angelo Garofalo\IEEEauthorrefmark{1}, Luca Valente\IEEEauthorrefmark{1},\\
 Suresh Sugumar\IEEEauthorrefmark{3}, Alessandro Giusti\IEEEauthorrefmark{2}, Davide Rossi\IEEEauthorrefmark{1}, Daniele Palossi\IEEEauthorrefmark{2}\IEEEauthorrefmark{4}
}
\IEEEauthorblockA{\IEEEauthorrefmark{1} Department of Electrical, Electronic and Information Engineering - University of Bologna, Italy}
 \IEEEauthorblockA{\IEEEauthorrefmark{2} Dalle Molle Institute for Artificial Intelligence - USI-SUPSI, Switzerland}
\IEEEauthorblockA{\IEEEauthorrefmark{3} Secure Systems Research Center - TII, United Arab Emirates}
 \IEEEauthorblockA{\IEEEauthorrefmark{4} Integrated Systems Laboratory - ETH Z\"urich, Switzerland}
 Contact author: maicol.ciani@unibo.it
}
\maketitle

\begin{abstract}
Autonomous Micro Aerial Vehicles (MAVs), with a form factor of \SI{10}{\centi\meter} in diameter, are an emerging technology thanks to the broad applicability enabled by their onboard intelligence. 
However, these platforms are strongly limited in the onboard power envelope for processing, i.e., less than a few hundred \SI{}{\milli\watt}, which confines the onboard processors to the class of simple microcontroller units (MCUs).
These MCUs lack advanced security features opening the way to a wide range of cyber-security vulnerabilities, from the communication between agents of the same fleet to the onboard execution of malicious code.
This work presents an open-source System-on-Chip (SoC) design that integrates a 64-bit Linux capable host processor accelerated by an 8-core 32-bit parallel programmable accelerator. 
The heterogeneous system architecture is coupled with a security enclave based on an open-source OpenTitan root of trust.
To demonstrate our design, we propose a use case where OpenTitan detects a security breach on the SoC aboard the MAV and drives its exclusive GPIOs to start a LED-blinking routine.
This procedure embodies an unconventional visual communication between two palm-sized MAVs: the receiver MAV classifies the sender’s LED state (on or off) with an onboard convolutional neural network running on the parallel accelerator; then, it reconstructs a high-level message in \SI{1.3}{\second}, 2.3$\times$ faster than current commercial solutions.
\end{abstract}

\input{01-Introduction}
\input{02-System_design}
\input{03-Security_UC}
\input{04-Use_case}
\input{05-Results}
\input{06-Conclusion}



\bstctlcite{IEEEexample:BSTcontrol}

\bibliographystyle{IEEEtran}
\bibliography{IEEEabrv,main}

\end{document}

%% file: 01-Introduction.tex
\section{Introduction} \label{sec:intro}

Autonomous Micro Aerial Vehicles (MAVs) are progressively gaining importance thanks to their ubiquitous sensing capabilities.
In the Internet of Things (IoT) ecosystem, nano-drones, i.e., palm-sized MAVs, can acquire and process information from different locations by flying where their presence is more important~\cite{lakshman2021integration}.
Therefore, they can exchange crucial data with fixed infrastructure or other drones, i.e., swarm operations.
Their miniaturized form factor enables a wide range of applicability, for example, in narrow spaces~\cite{quadrotor_nanodrone} and human surroundings~\cite{blimp1_nanodrone}, but it limits the class of processors they can host aboard.
This \textit{i}) lower-bounds the computational/memory complexity of the algorithms that can run aboard and \textit{ii}) forces the main drone's mission computer to simple microcontroller units (MCUs) that lack advanced cyber-security features.

In this emerging new era of connected and collaborating IoT devices/nano-drones, reliable security and privacy mechanisms are needed to protect assets and data collected or generated~\cite{hwang2015iot}.
The security cornerstone of IoT devices is the Root of Trust (RoT), where critical assets are kept isolated and protected, the code executed is authenticated, and its integrity is verified~\cite{RoT}.
Most modern IoT devices rely on hardware to ensure their RoT and therefore build the whole security stack on top of it, following the \textit{chain of trust} principle~\cite{RoT}.
Despite RoTs provide a solid hardware/software security foundation, there are several types of attacks potentially compromising the drones' operations, such as man-in-the-middle, denial of service, spoofing, jamming, rogue data injection, routing attack, etc.~\cite{Yahuza2021iod}.

Current Commercial Off-The-Shelf (COTS) nano-drones platforms, such as the Bitcraze Crazyflie typically host low-power 32-bit MCUs such as the STM32F4 as main mission computer~\cite{quadrotor_nanodrone}.
This class of MCUs provides sufficient computing power to guarantee basic functionalities such as low-level control loops, state estimation, and cryptographic encoding.
Although they lack both a security enclave and RoT; therefore, they can not guarantee hardware isolation of code execution or support full-fledged operating systems capable of software isolation of different parts of the applications running on them.
Similarly, more computationally-capable SoCs for nano-drones, such as the GWT GAP8 processors~\cite{GAP8} available as a companion board for the Crazyflie nano-drone, i.e., AI-deck, still lacks RoT and security enclave able to take control of the whole system in case of attacks.
In this work, we present an open-source SoC design of a mission computer for autonomous nano-drones, which includes silicon secure enclave and RoT by integrating the OpenTitan reference design~\footnote{\url{https://opentitan.org/}}. 

\begin{figure*}[t!]
\centerline{\includegraphics[width=\textwidth]{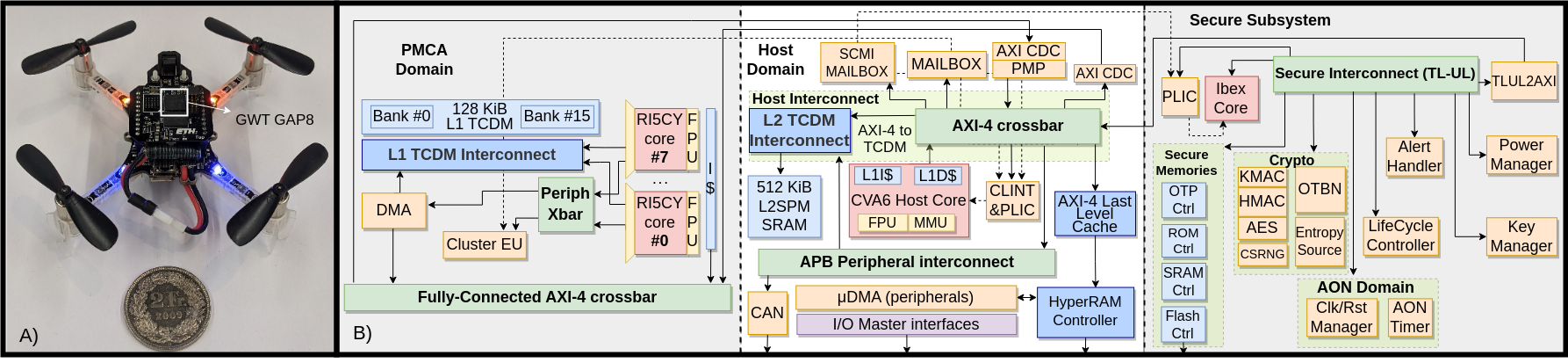}}
\caption{A) The Bitcraze Crazyflie equipped with the GWT GAP8 SoC. B) The proposed SoC architecture envisioned as alternative MCU aboard the nano-drone.}
\label{fig:soc}
\end{figure*}

The SoC is built around a 64-bit RISC-V CVA6 core featuring full Linux support and an 8-core cluster of 32-bit RISC-V cores acting as a software-programmable accelerator enabling vision-based tasks.
Our work uses and enhances OpenTitan, the first collaborative open-source RISC-V-based silicon RoT, to support service request handling through an System Control and Management Interface (SCMI)~\footnote{\url{https://developer.arm.com/documentation/den0056/d/?lang=en}} mailbox, master on the host domain, and secure GPIO handling connected to LEDs.

We showcase our system design with a novel and field-proven use case of \textit{Unconventional Visual Communication} (UVC) between nano-drones exchanging messages by LED blinking.
In the context of visible light communication, machine learning techniques are used on the receiving end to implement signal demodulation~\cite{9352471}, to recover the modulated signal from rolling-shutter images~\cite{6477759,Hsu:20}, or to find locations of transmitters in images, before analyzing those regions of interest separately~\cite{9700576,9780193,trajectories}. 
In contrast, our approach feeds raw images directly to a CNN to directly extract binary signal information (LEDs on or off), independently of the relative location of the drone transmitting the signal.

Once a cyber-attack compromises a nano-drone in the fleet, the radio channel cannot be trusted and the UVC is triggered, which depends exclusively on the secure OpenTitan sub-module and results in an SOS message emitted by the blinking LEDs.
We show how other nano-drones equipped with the same SoC can reconstruct the SOS message by analyzing a video stream.
By running a convolutional neural network (CNN), we assess the LED state of each input image.
Then, a simple state machine continuously analyzes the series produced by the CNN and retrieve any custom message, such as the SOS one. 

Our main contribution is the development of a novel SoC for drones' autonomous navigation providing cyber-security features which are keys in the proposed UVC-based use case.
In detail: \textit{i}) we integrate the OpenTitan secure subsystem into the navigation controller SoC; \textit{ii}) we develop and field-test a simple light-based communication between multiple nano-drones in the swarm.
With a power envelope of \SI{250}{\milli\watt} and a silicon footprint of \SI{9}{\square\milli\meter}, the proposed SoC can recognize an SOS message in \SI{1.3}{\second} performing 2.3$\times$ faster than a Crazyflie nano-drone equipped with an AI-deck, while offering support for a security enclave and full-fledged operating system.

%% file: 02-System_design.tex
\section{System Architecture} \label{sec:system}

This section presents the SoC architecture in Figure~\ref{fig:soc}.
It consists of a heterogeneous system architecture composed of a 64-bit application processor implementing flight control functions as well as auxiliary functions such as network stack, a parallel programmable accelerator for mission control functions, and a secure enclave based on OpenTitan IPs.

The SoC is built around the CVA6 core, a 6-stages, single-issue, in-order, 64-bit RISC-V core supporting the RV64GC ISA variant, virtual memory, three execution privilege levels, physical memory protection (PMP), and is capable of booting the Linux OS.
CVA6 has 16KB of L1 I-cache and 32KB of write-through L1 D-cache, which enable simple coherency with other masters to the crossbar interconnect, which implements high-bandwidth, low-latency 64-bit AXI4 protocol.
The \textit{host domain} contains a scratchpad memory (L2SPM) and a complete set of peripherals such as I2C, (Q)SPI, CPI, SDIO, UART, CAN, PWM, I2S.
Moreover, the host embeds also a standard Platform Level Interrupt Controller (PLIC), a Core Local Interrupt (CLINT), a controller for Cypress Semiconductor's external HyperRAM memories, and a Last Level Cache (LLC) to filter accesses to the external HyperRAM memory improving system performance.
Peripheral data is transferred from/to the scratchpad memory through a dedicated DMA, called $\mu$DMA.

The Programmable Multi-Core Accelerator (PMCA) of the system is built around 8 CV32E-based processors which share 16×8KB SRAM banks (128KB L1SPM).
The cores implement RV32 extension with many ML and DSP features such as hardware loops, MAC\&Load operation, SIMD operations, and post-increment LD/ST.
With SIMD, the operands’ width can be reduced to double or quadruple the number of operations per cycle.
The cluster also implements FPUs supporting FP32 and FP16 with SIMD support and features a two-level I-cache (512B for each core and 4KB shared) to speed-up execution of data-parallel tasks typical of drone mission control functions and deep neural networks for objects and pattern recognition.
The architecture of the cluster is optimized for ML algorithms in embedded applications: it exploits scratchpad memories with DMA access, double buffering and custom ISA extensions to optimize memory utilization and computation.

The third key component of the drone navigation SoC is the secure enclave based on the OpenTitan architecture, acting as an on-chip Root of Trust (RoT), providing security services.
The Ibex core is the main processor and is in charge to orchestrate the secure boot and all the RoT functionalities.
Ibex controls four main components.
The AON domain includes power, reset, clock management and a timer.
The secure memories module includes One Time Programmable (OTP) memories which store security keys and seeds.
The crypto module includes specialized accelerators such as Advanced Encryption Standard (AES), Hashing (HMAC and KMAC) and Big Number Accelerator (OTBN) for Rivest–Shamir–Adleman (RSA) and Elliptic Curve Cryptography (ECC).
The security module includes key manager, life cycle controller and alert handler.

The host domain requires to exploit the hardware cryptography accelerators of the secure subsystem but it must not access directly its internal memory map for security reasons.
Instead, it can only encode commands writing into a dedicated mailbox compliant with ARM System Control and Management Interface (SCMI).
The host domain populates the shared memory of the mailbox and then it raises an interrupt to the secure subsystem's core through a dedicated memory mapped register.
The Ibex core reads the content of the mailbox and executes the command encoded in it.
At the end of the execution, the Ibex core raises back another interrupt to the host domain's core.
Moreover, OpenTitan has its own internal timer that can trigger periodic interrupts.
In this way the Ibex core can be periodically waken up in order to perform anomaly detection checks by analyzing the content of CVA6 and cluster's memories as well as external peripherals.

%% file: 03-Security_UC.tex
\section{Security Use Case} \label{sec:sec_usecase}

Our use case envisions multiple nano-drones cooperatively operating and exchanging periodic data (e.g., mission commands, etc.) via radio (e.g., WiFi, BTLE, etc.).
In this scenario, we address the following two \textit{threat models}.

\begin{figure}[t]
  \includegraphics[width=\columnwidth]{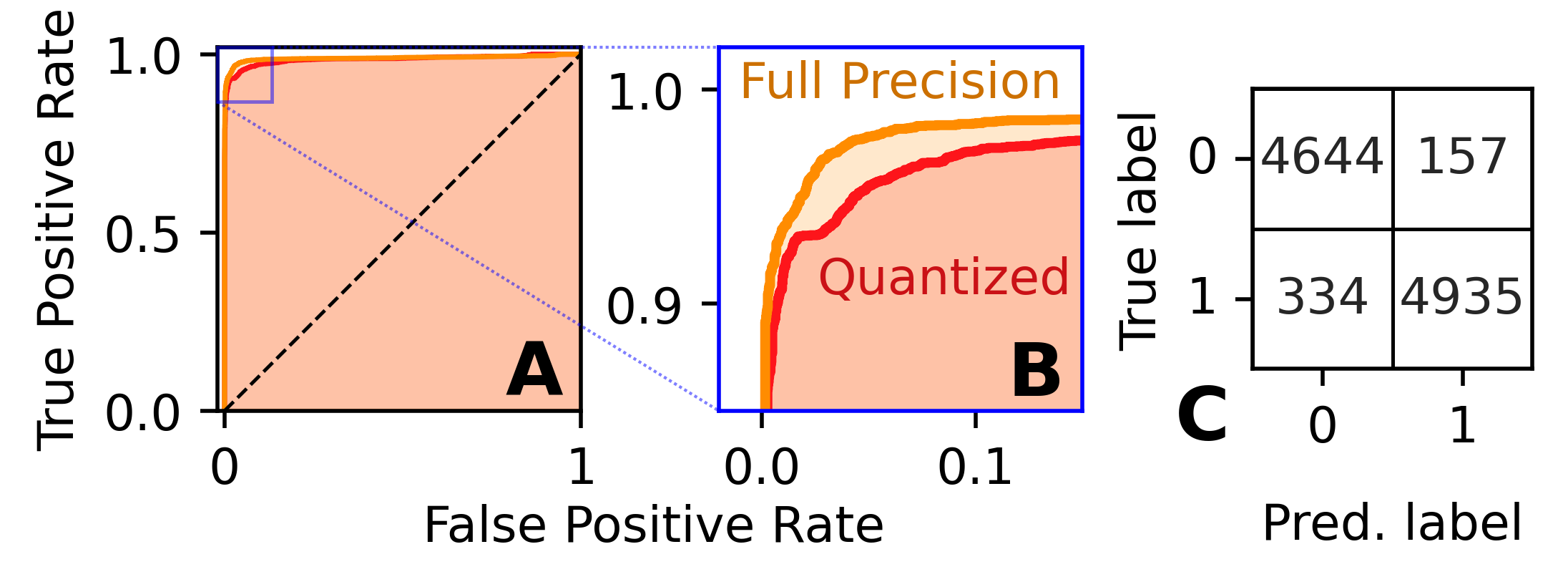}
  \caption{Performance on the testing set. \textbf{A)} ROC Curves for full-precision and quantized models. \textbf{B)} Zoom-in of A. \textbf{C)} Confusion matrix.}
  \label{fig:roc-curve}
\end{figure}

\textbf{Man-In-The-Middle.}
A man-in-the-middle attack enables the attacker to intercept the communication and exchange malicious data with the drones.
Following the zero trust policy~\cite{rose2020zero}, where we always authenticate and never trust, the drones periodically check that the received data are original and transmitted by an authenticated fleet member.
If the authenticity cannot be verified, OpenTitan assumes that both the communication radio channel and the rest of the SoC are potentially compromised.
Therefore, to notify the rest of the fleet about this situation, it enables the UVC blinking procedure by driving its secure GPIOs (exclusively connected to the secure subsystem) to transmit an informative SOS message.
Other fleet drones -- in line-of-sight with the transmitter one -- can simultaneously or alternatively monitor peers' activity, distributing and time-interleaving the computational overhead for the message decoding.

\textbf{Anomaly/Intrusion Detection.}
For this use case, we assume that there is a minimal anomaly/intrusion detection mechanism~\cite{Galvan2021anomaly,lunardi2022arcade} running on the Ibex core of OpenTitan, which is a secure region by construction.
Since OpenTitan is the master of the TLUL-to-AXI interface on the host AXI-4 crossbar, it can monitor the activity of sensors (e.g., accelerometer, cameras, etc.).
Then, if OpenTitan detects an anomaly, it can assume that the host domain, including the communication links, is compromised and triggers the UVC procedure.
The specific implementation of a detection mechanism is out of the scope of this paper.
Instead, we focus our work on the OpenTitan integration/isolation from the rest of the SoC and the implementation of the attack response method, i.e., the UVC exchanging SOS messages by demonstrating it with two Crazyflie nano-drones equipped with the AI-deck.

%% file: 04-Use_case.tex
\section{Unconventional Visual Communication} \label{sec:use_case}

\begin{figure}[t]
  \includegraphics[width=\columnwidth]{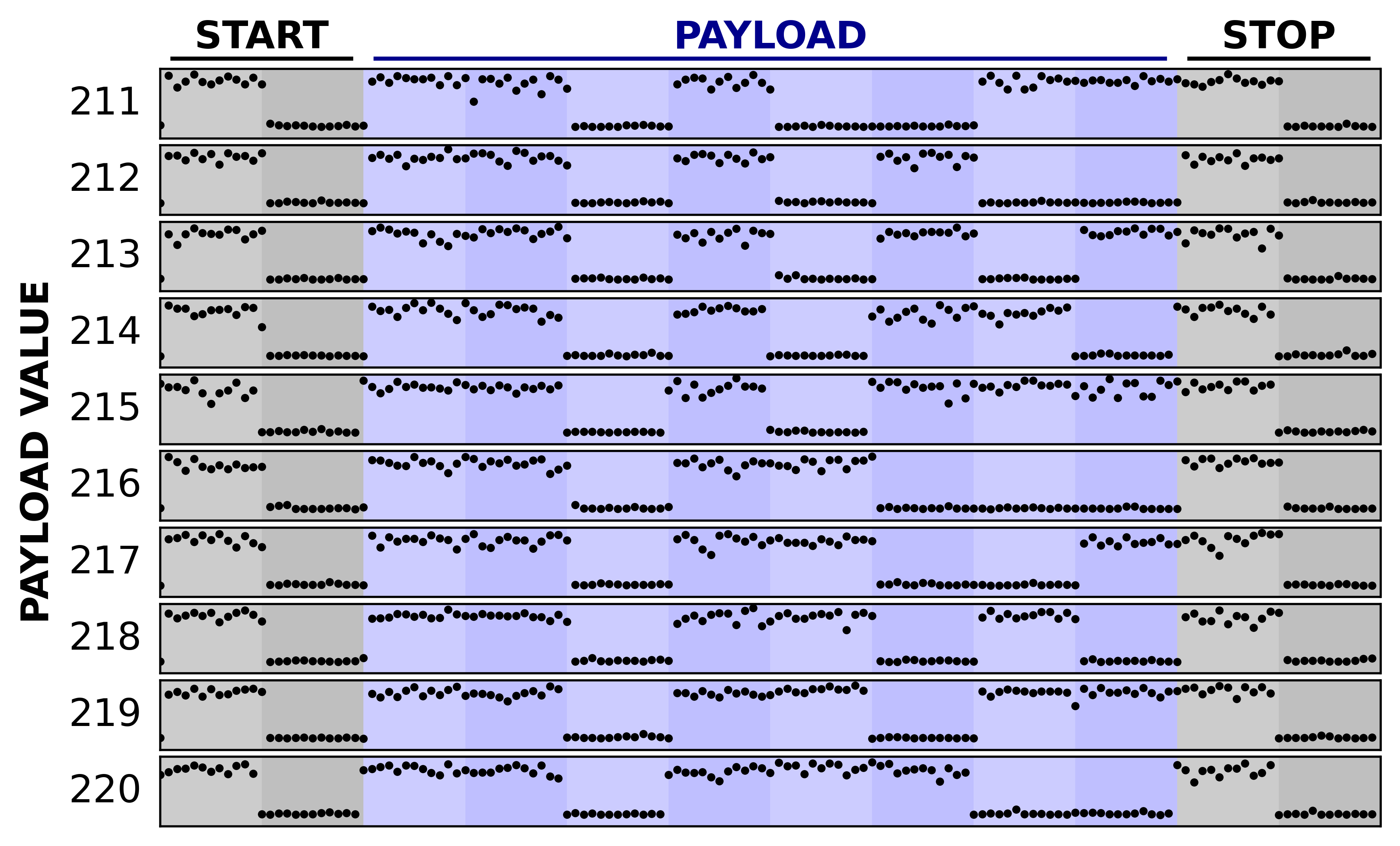}
  \caption{Small dots denote individual CNN predictions of the transmitter LED state (12 per bit), before averaging. Vertical shaded areas denote the bit clock. Colors denote 2 start, 8 payload and 2 stop bits.}
  \label{fig:incremental-samples}
\end{figure}

\textbf{CNN training.} 
To decode the message transmitted by OpenTitan of a compromised drone, we use a lightweight CNN with the field-proven architecture of PULP-Frontnet~\cite{frontnet}.
The model input is a $160\times96$ pixels grayscale image; the output estimates the state of the LEDs of a Crazyflie nano-drone that is assumed to be visible in the image (either all on or all off), regardless of the drone position in the frame.
Datasets are acquired with a monochrome QVGA camera from an \textit{observer} drone facing a \textit{transmitter} drone, which toggles its four LEDs every $\sim\SI{0.4}{\second}$.

The dataset is collected in a room equipped with a 18-camera Optitrack motion capture system, which tracks both drones.
The transmitter drone flies in the central part of the room, while the observer records images while following a circular path around the transmitter.  
In this way, we maximize the variability of the images' backgrounds.  
The transmitter drone is automatically controlled in order to: \textit{i}) always stay within the field of view of the observer; \textit{ii}) move to random positions uniformly distributed on the observer's image plane; and \textit{iii}) always lie at a distance of \SIrange{0.2}{1.8}{\meter} from the observer.  
The dataset is composed of 36 flights, where frames are split to build a training set (the first 72\%), a testing set (the intermediate 19\%), and a validation set (the last 9\%).
A 10-frame gap is ignored between different sets in the same flight, to ensure that no similar frames appear in different sets.  
This results in \SI{37}{\kilo\nothing} frames for training, \SI{10}{\kilo\nothing} frames for testing and \SI{5}{\kilo\nothing} frames for validation; each frame is labeled with the corresponding ground-truth LED state; the two states are equally represented in all sets.

\textbf{Message encoding and decoding.}
Messages with an 8-bit payload are encoded to a simple self-clocking binary line protocol~\cite{halsall1995data} that produces 12-bit packets, including 2 start bits and 2 stop bits.
On our prototype, we employ the 8-cores GAP8 SoC manufactured in TSMC \SI{55}{\nano\meter} technology capable of \SI{22.65}{\giga Op/\second} at \SI{4.24}{\milli\watt/\giga Op}.
With this SoC, the bit stream is transmitted by modulating the LED state at a rate of 2.5 bits per second; each 12-bit packet is therefore transmitted in \SI{4.8}{\second}.
The observer drone acquires images at 30 frames per second (FPS).
Each image is fed to the CNN, which estimates the LED's state in each frame.
Each bit appears in 12 consecutive frames.
First, the \textit{bit clock} is determined from this sequence, then each bit in the bit stream is estimated by averaging the corresponding 12 CNN outputs and thresholding the result.  
Messages are decoded from the bit stream starting with a reserved start flag. 
If necessary, error detection and correction codes~\cite{hamming1950error} can be implemented on top of this approach.

%% file: 05-Results.tex
\section{Experimental results} \label{sec:results}

\subsection{Deployment on Crazyflie with AI-deck}

Figure~\ref{fig:roc-curve} reports the CNN performance on the testing set; the full-precision model achieves an Area Under the ROC curve score of 98.88\%, with negligible performance loss ($-$0.09\%) after 8-bit integer quantization.
After binarizing outputs at a threshold of 0.5, the model achieves an accuracy of 95.1\%.

\begin{table}[t]
    \centering
    \caption{Power consumption and Area occupation}
    \begin{tabular}{|c|c|c|c|c|c|} \hline
                           & Area      & Leakage & Dynamic            & Max Freq & Max Power \\ 
                          & ($mm^2$)   & ($mW$)  & ($\frac{uW}{MHz}$) & ($MHz$)     &  ($mW$) \\ \hline
         Top              & 7.28       & 4.23    & 214.7              & 450       & 100.53 \\\hline
         CVA6             & 0.49       & 4.79    & 47.5               & 900       & 47.54  \\ \hline
         PMCA             & 1.56       & 5.78    & 206                & 400       & 88.18  \\ \hline
         Mem Ctrl.        & 0.27       & 0.14    & 2.3                & 450       & 1.16   \\ \hline
         Opentitan        & 0.86       & 4.53    & 16                & 350       & 10.13   \\ \hline
         Total            & 7.28       & 19.47   & 486.5             & -         & 247.54  \\ \hline
    \end{tabular}
    \label{tab:power_table}
\end{table}

The end-to-end message transmission is assessed with an experiment in which the transmitter drone sends a sequence of 256 messages with a payload values from 0x00 to 0xFF. 
The observer drone, placed at a fixed distance of \SI{30}{\centi\meter}, is always in line-of-sight with the transmitter one and decodes the received messages, at 30 FPS, with the quantized CNN.
All 256 messages are decoded correctly. 
Figure~\ref{fig:incremental-samples} reports a subsequence of the received messages and a supplementary video demonstration is provided at \url{https://youtu.be/TClcuUWJe0U}.

\subsection{Physical Implementation \& Performance Evaluation}

\begin{figure}[t!]
  \centering\includegraphics[width=\columnwidth]{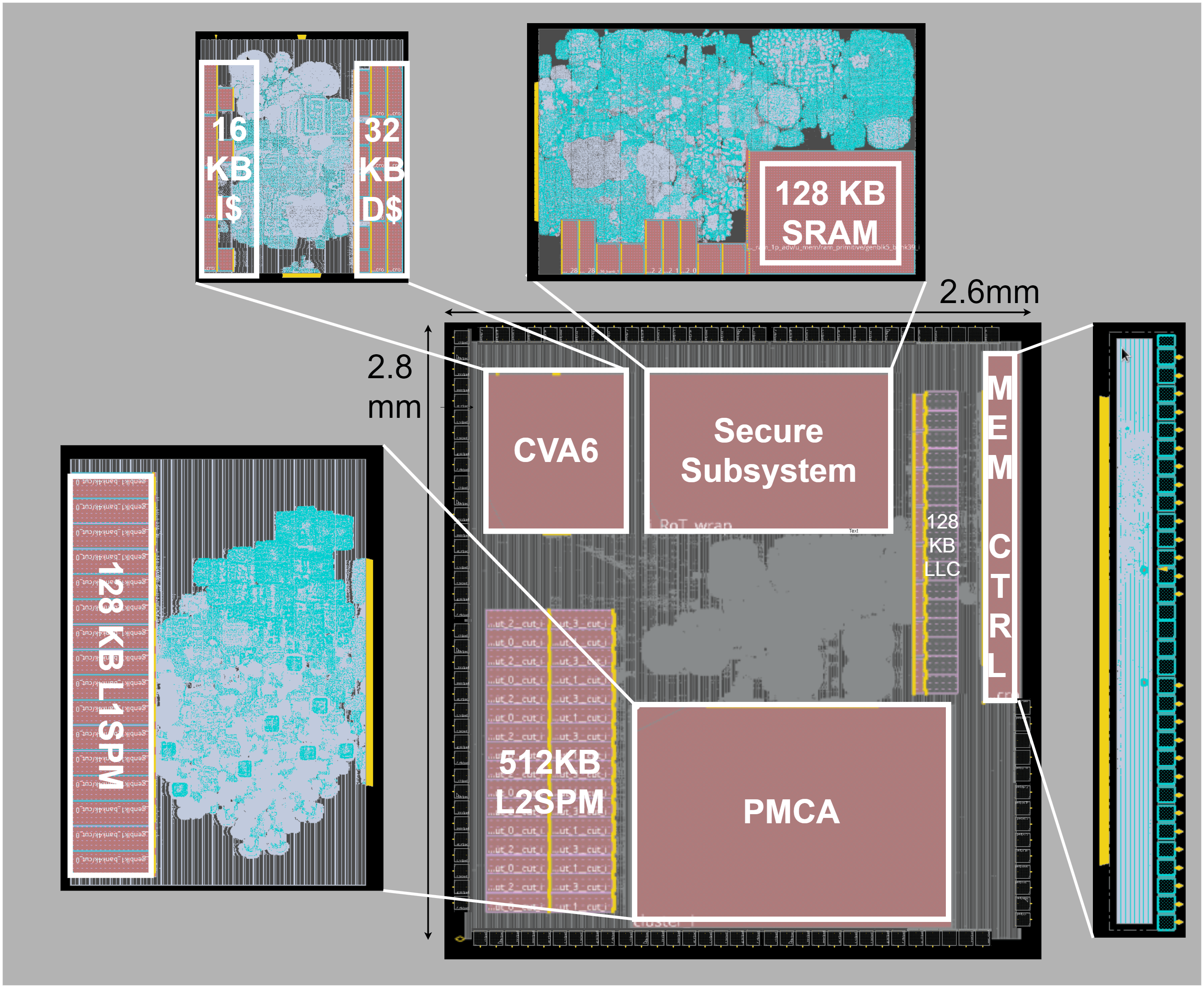}
  \caption{In the middle, the layout of the SoC. Around it, the layouts of the PMCA, secure subsystem, HyperBus and CVA6.}
  \label{fig:layout}
\end{figure}

The proposed SoC has been implemented in the Global Foundries \SI{22}{\nano\meter} FDX technology, employing the Synopsys Design Compiler for the logical synthesis and the place and route with Cadence Innovus. 
For the SoC's signoff we used the Synopsys PrimeTime, considering the worst case operating corner for a nominal supply voltage of \SI{0.8}{\volt} (SS, \SI{0.72}{\volt}, 125\textdegree C/-40\textdegree C), while power analysis was performed in typical operating conditions (TT, \SI{0.8}{\volt}, 25\textdegree C)
The layouts of the SoC and the main subsystems composing it are shown in Figure~\ref{fig:layout}, while Table~\ref{tab:power_table} summarizes the physical implementation.

To evaluate the performance of the proposed SoC on the UVC use-case, we first deploy our CNN on the PMCA of GAP8 SoC hosted on the COTS Carzyflie nano-drone.
The full inference of the DNN on the proposed SoC takes \SI{3.7}{\mega cycles}, meaning that each payload's bit can be recognized by the receiver drone in \SI{126}{\milli\second} and that a full message described in Figure~\ref{fig:incremental-samples} can be recognized in \SI{1.3}{\second} by our proposed SoC.
This is 2.3$\times$ faster than the same application running on the GAP8 SoC (@\SI{175}{\mega\hertz}).
By coupling a linux-capable application core with a parallel programmable accelerator and a secure enclave within the power budget of \SI{250}{\milli\watt} and a footprint of \SI{9}{\square\milli\meter}, our SoC represents an appealing solution for secure and high-performance mission computers for nano-drones, paving the way for a wide range of new secure applications.

%% file: 06-Conclusion.tex
\section{Conclusion} \label{sec:conclusion}

We present an open-source SoC design for ultra-low power mission computers compatible with the limited power envelope of nano-UAVs.
Our design provides sufficient computational resources to enable autonomous navigation tasks while enabling advanced hardware security such as RoT.
The SoC is built around a 64-bit RISC-V CVA6 core featuring full support for Linux accelerated by an 8-core cluster of 32-bit RISC-V cores acting as a software-programmable accelerator for mission control tasks.. 
We integrate a security enclave based on an open-source RoT, enabling a multi-drone UVC use case.
In our scenario, OpenTitan detects a security breach on the SoC and communicates an SOS message to a receiver drone by sending it through LEDs blinking.
The receiver nano-drone can detect the message by running on the programmable accelerator a visual CNN and a simple state machine that decodes it.
With a power envelope of \SI{250}{\milli\watt} and a silicon footprint of \SI{9}{\square\milli\meter}, the proposed SoC can recognize an SOS message in \SI{1.3}{\second} performing 2.3$\times$ faster than a COTS baseline equipped with the GAP8 SoC, while offering support for a security enclave and full-fledged operating system.